\documentclass[useAMS,usenatbib]{mn2e}
\usepackage{psfig}
\title[Chaos in MRI-driven Turbulence]{Chaos in Turbulence Driven by 
the Magnetorotational Instability}

\author[Winters, Balbus, \& Hawley]
  {Wayne F. Winters, Steven A. Balbus, and John F. Hawley \\
Dept. of Astronomy, University of Virginia, PO Box
3818, Charlottesville, VA 22903, USA.  }
\date{Accepted.  Received.}
\pagerange{\pageref{firstpage}--\pageref{lastpage}} \pubyear{2002}

\begin{document}
\maketitle
\label{firstpage}

\begin{abstract}

Chaotic flow is studied in a series of numerical magnetohydrodynamical
simulations that use the shearing box formalism.  This mimics important
features of local accretion disk dynamics.  The magnetorotational
instability gives rise to flow turbulence, and quantitative chaos
parameters, such as the largest Lyapunov exponent, can be measured.
Linear growth rates appear in these exponents even when the flow is
fully turbulent.  The extreme sensitivity to initial conditions
associated with chaotic flows has practical implications, the most
important of which is that hundreds of orbital times are needed to
extract a meaningful average for the stress.  If the evolution time in
a disk is less than this, the classical $\alpha$ formalism will break
down.
\end{abstract}


\section{Introduction}

Astrophysical accretion disks are able to evolve because angular
momentum is extracted from fluid elements and transported outward.
This is effected by the presence of nonvanishing radial-azimuthal
components of the Maxwell and Reynolds stress tensors, produced by
magnetohydrodynamic (MHD) turbulence driven by the magnetorotational
instability (MRI) \citep{bh98}.  As there is no analytic theory of MHD
turbulence at hand (nor is there one in sight), large-scale numerical
simulation has been the main avenue of progress toward understanding
its properties.

Many numerical studies make use of the local ``shearing box''
approximation \citep{hgb95}.  The shearing box is an invaluable tool
for studying local flow dynamics in detail, and for resolving turbulent
flow with the largest possible dynamical range.  MHD turbulence, like
all true turbulence, should be chaotic, as quantified formally by a
measured positive Lyapunov exponent.  What is less clear, but of great
astrophysical significance, is whether long term flow averages are even
well-defined.  To put the question most starkly, imagine macroscopically
identical disks, with fluid perturbations that vary by a tiny amount.
The fine scale description of their internal turbulence will surely
differ, but will quantities such as the stress tensor components converge
to the same values in the long term?  This question goes directly to the
heart of the standard $\alpha$ disk formalism \citep{ss73} which assumes
that disk transport may be described by a spatially constant or slowly
varying $\alpha$ parameter.  (The quantity $\alpha$ is defined as the
ratio of radial-azimuthal component of the stress tensor $T_{R\phi}$
to the gas, or gas plus radiation, pressure.)  How well supported is
this assumption?

This paper begins to study these complex issues by examining the
chaotic nature of the MHD turbulence.  The presentation is organized as
follows.  In \S 2 we briefly review the shearing box.  In \S3 we
present the results of experiments designed to reveal how the measured
properties of turbulent flow are related to both computational and
physical input parameters.  In \S4, we carry out a series of
experiments that demonstrate qualitatively that the MHD turbulence is
chaotic, and in \S5 this is quantified by computing the Lyapunov
exponents for a set of simulations.  Our conclusions are summarized in
\S6.

\section{Shearing Box System}

The shearing box system, described in \cite{hgb95}, is designed to
represent a very local section of an accretion disk, viewed in 
corotating coordinates with angular frequency $\Omega$.  Starting with
the full set of dynamical equations in cylindrical coordinates, the
equations are locally expanded about a fiducial cylindrical radius $R$,
with $(dx, dy, dz)$ corresponding to cylindrical coordinates $(dR,
Rd\phi, dz)$.  The computational domain is then a Cartesian box, but
with the rotational inertial forces (Coriolis and centrifugal) retained.
The centrifugal force nearly balances gravity, leaving a remnant tidal
force linear in $x$.   All other forces are directly retained.
For this system, the ideal MHD equations of motion
become,\hfil\vfil\eject
\begin{eqnarray}\label{euler}
\frac{\partial {\mathbf v}}{\partial t} +
\left({\mathbf v}\cdot \nabla \right){\mathbf v} =
- \frac{1}{\rho}\nabla \left(P + \frac{B^{2}}{8\pi}\right) +
\frac{\left({\mathbf B}\cdot \nabla \right){\mathbf B}}{4\pi \rho}
\nonumber\\
- 2{\mathbf \Omega}\times {\mathbf v} +
2q\Omega^{2}x\mathbf{\hat{x}} -
\Omega^{2}z\mathbf{\hat{z}}.
\end{eqnarray}
The mass conservation equation 
\begin{equation}
\frac{\partial \rho}{\partial t} + \nabla \cdot \left( \rho {\mathbf v}
\right) = 0,
\end{equation}
the adiabatic internal energy equation,
\begin{equation}
\frac{\partial \rho \epsilon}{\partial t} +
\nabla \cdot \left(\rho \epsilon {\mathbf v}\right)+
P\nabla \cdot {\mathbf v} = 0,
\end{equation}
and the induction equation
\begin{equation}
\frac{\partial {\mathbf B}}{\partial t} =
\nabla \times \left({\mathbf v}\times {\mathbf B}\right),
\end{equation}
retain their usual forms.  
The internal energy per particle $\epsilon$ is defined by 
\begin{equation}
P = \rho \epsilon \left(\gamma - 1\right). 
\end{equation}
For an isothermal gas, the energy equation is replaced with
\begin{equation}
P \rho^{-1} = \mathrm{constant}.
\end{equation}
The variables in these equations have their usual meanings.  We shall
also introduce the constant $q$, which is a parameter describing the
local radial dependence of the angular frequency, i.e.,  $q = - {d \ln
\Omega}/{d \ln R}$.  For a Keplerian angular momentum distribution, $q
= 3/2$.  In this study we have ignored vertical gravity, and have thus
dropped the $\Omega^2 z \mathbf{\hat{z}}$ term from equation
(\ref{euler}).

These equations are solved using the time-explicit, operator-split
finite differencing algorithm of the ZEUS code for hydrodynamics
\citep{sn92a} and MHD \citep{sn92b, hs95}. 
The shearing box computational domain extends in $x$, $y$, and $z$
respectively
from ${-L_{x}}/{2}$ to $+{L_{x}}/{2}$, 0 to $L_{y}$, 0 to $L_{z}$.  The
boundary conditions are periodic in both the $y$ and $z$ directions, and
``shearing periodic'' in $x$, meaning we equate
\begin{equation}
f(x,y,z)=f(x+L_{x},y-q\Omega L_{x}t,z).
\end{equation}
The azimuthal velocity has an additional correction to
offset the angular velocity difference between the inner and outer
radial boundaries: 
\begin{equation}
v_{y}(x,y,z) = v_{y}(x+L_{x},y-q\Omega L_{x}t,z)+q\Omega L_{x}.
\end{equation}

The initial state is one of uniform density and pressure, and the
velocity profile is ${\mathbf v} = -q\Omega x \mathbf{\hat{y}}$.  In this
paper, our initial magnetic field configurations are either a uniform
toroidal magnetic field, or a vertical magnetic field varying
sinusoidally in the radial $x$ direction.  The magnetic field strength
is  described by the standard plasma $\beta$ parameter
\begin{equation}\label{beta} \beta \equiv {8\pi P\over B^2},
\end{equation} the ratio of the gas to magnetic pressures.  The MRI is
triggered by seeding the initial equilibrium state with small
pressure and velocity fluctuations. 

\section{Saturation Amplitude in the Shearing Box: a Survey}

One of the goals of shearing box simulations has been to try to
understand what determines the sustained saturation amplitude of the
MRI-driven turbulence.  In accretion disk applications, the focus is
often on the $T_{r\phi}$ component of the combined Reynolds and Maxwell
stress tensor, 
\begin{equation} T_{r\phi} = \left( - {B_{x} B_{y}}/{4
\pi}+\rho v_{x} \delta v_{y}\right). 
\end{equation}
Here $\delta v_{y}$ is the azimuthal velocity fluctuation, obtained by
subtracting the equilibrium shear flow.  This can be directly measured
in shearing box simulations.  The volume-averaged stress is usually
normalized to the averaged gas pressure, a quantity equivalent to the
\citep{ss73} $\alpha$ parameter.

Even with a system as simple as the shearing box, there are many
possible factors, both computational as well as physical, that could
in principle govern the saturation amplitude.  These
include the initial field strength and geometry, box size and aspect
ratio, value of $\Omega$, background gas pressure, and numerical
resolution.  Many of these are physically significant only
in relation to others.  For example, $\Omega$ is an arbitrary time
scale in the system.  The shearing box system is independent of
$\Omega$ so long as $q$ remains constant and the physical box size
changes proportionately.   The formal scaling invariance is that if
\begin{equation}
\Omega \rightarrow c \Omega,
\end{equation}
where $c$ is an arbitrary constant, then the equations remain unchanged
if 
\begin{equation}
(x, t) \rightarrow (x/c, t/c).
\end{equation}
We will make use of this fundamental property in \S4 when we explore chaotic
behavior. 

For the cases of an initial uniform field oriented along either the $y$
or $z$ axes, \cite{hgb95} found that the saturation amplitude of the
turbulence depended upon both the box size and initial field strength.
When the field had a vanishing mean value over the computational box
volume, the final outcome was much less sensitive to the initial field
energy.  Here, we present a series of zero mean field simulations
designed to test how the box size affects the overall level of the
turbulence.  The baseline calculation has $q = 1.5$ corresponding to a
Keplerian background, and a box size of $L_x =1$, $L_y= 2\pi$, and
$L_z=1$, set so that $L_z = c_s^2/\Omega$ where $c_s^2=P/\rho$.  The
actual values used are $P=10^{-6}$ and $\Omega = 0.001$.  The gas is
adiabatic with $\gamma = 5/3$.  The initial field is vertical and
varies sinusoidally in the $x$ direction, $B_{z} \propto
\sin(x/L_x)$.  The field strength set so that the volume average
magnetic energy corresponds to $\beta = 800$.  The maximum vertical field
has $\beta=400$, which gives a fastest growing wavelength for the MRI of
$\lambda_{\rm max} = 8\pi/15^{1/2} (v_A/\Omega) = 0.46$.
The grid resolution is
$32 \times 64 \times 32$ zones in $(x,y,z)$.  This setup corresponds to
the so-called ``fiducial box'' used in \cite{hgb95, hgb96} and
\cite{betal95}.  

\begin{figure}
\centerline{
\psfig{file=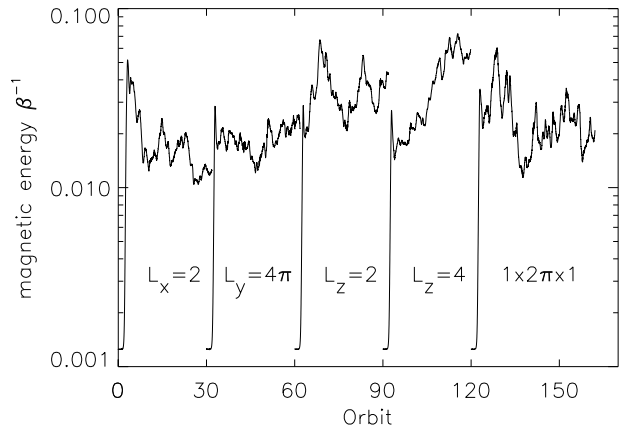,width=3.5in}}
\caption{Magnetic energy histories for an ensemble of shearing box runs
with different box sizes.  Magnetic energy is expressed in terms of the
ratio of the magnetic pressure to initial gas pressure, $\beta^{-1}$.
The zero point in time for each run is offset
by 30 orbits for ease of comparison.  Each curve is labeled by the
dimension that is increased.  The fiducial box is the rightmost plot.
\label{zadiab}}
\end{figure}

In addition to the fiducial run, we computed four additional
simulations with different box dimensions.  In three simulations we
doubled each of the box dimensions in turn, and in the final simulation
of this set we quadrupled the vertical size.  To maintain a constant
resolution number of grid zones per unit length, the number of grid
zones used was increased appropriately.  Figure \ref{zadiab} shows the
magnetic energy per unit volume evolving over 30 orbits for each run in
this series of models.  While one must be cautious extrapolating from
this limited time baseline, the results generally suggest that increasing
the vertical dimension is the most important effect over the long term.
The temporal variations in magnetic energy within a given run are, however,
quite large, often well in excess of the running average.

\begin{figure}
\centerline{
\psfig{file=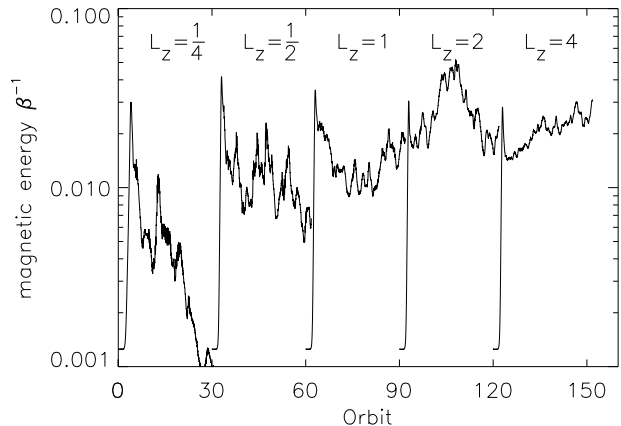,width=3.5in}}
\caption{Magnetic energy histories for an ensemble of shearing box runs
with different vertical box sizes.
Magnetic energy is expressed in terms of the
ratio of the magnetic pressure to initial gas pressure, $\beta^{-1}$.
The zero point in time for each run is offset
by 30 orbits for ease of comparison.  Each curve is labeled by the
value of $L_z$.
\label{ziso}}
\end{figure}

To isolate the effect of the vertical size, we carried out a second set
of simulations in which $L_z$ is varied, from $L_z = 1/4$ to $L_z=4$,
and all other factors are held fixed, including the grid zone size
$\Delta z$.  The initial magnetic field is as described above.  Here we
used an isothermal equation of state, $\gamma = 1.0$ which keeps the box
from heating due to dissipation of the turbulence.  Figure \ref{ziso}
shows the time evolution of the magnetic energy and even over a limited
time baseline clearly reveals a general trend toward increasing magnetic
energy with increasing $z$ box size, again with significant variation.
The most striking conclusion is that there is a lower bound to the
vertical size required for amplification to occur:  the field energy
declines dramatically in the $L_z = 1/4$ model, and even the initial
linear growth rate is reduced.  The fastest growing vertical wavelengths
for initial field strengths with $\beta < 1350$ no longer fit inside
the box, and evidently losses win over the reduced power input at the
top of the turbulent cascade.

For box sizes $L_z=1/2$ and larger the turbulence is sustained over the
evolution period of 30 orbits.  However, other trends are visible.
First, the amplitude of the initial spike of linear growth saturation
decreases with increasing $L_z$.  Second, the amplitude of the magnetic
energy variations in the turbulent state is reduced with increasing box
size.

These history plots demonstrate that it is not a simple matter to
quantify the saturation amplitude of the turbulence in a shearing box,
or to determine unambiguously the influence of physical parameters in
setting that amplitude.  It is not obvious, in general, over what
length of time one should average to obtain a characteristic amplitude;
nor is it even obvious that such a baseline must always exist.  Large
fluctuations, even in volume-averaged data, are the rule.  This is the
overarching and defining characteristic of the MHD turbulence:  it is
chaotic.

\section{Onset of Chaos}

It is clear that changing the physical parameters of a shearing box run
can result in different levels of turbulence.  Here we show that almost
exactly the {\it same} physical parameters, but with tiny variations in
numerical value, leads to macroscopically different levels of
turbulence.  The sensitivity of the system to infinitesimal
perturbations is a key feature of chaos.

The first set of simulations consists of three shearing boxes with
different $\Omega$ values, but which have been rescaled so as to be
physically equivalent.  All three have $q = 1.5$, $\beta = 800$,
$\gamma = 1.0$, $B_{z} \propto \sin (x/L_x)$, and $32 \times 64
\times 32$ grid zones.  One run uses the standard box size of $1\times
2\pi \times 1$ with $\Omega = 0.001$.  The other two have $\Omega =
0.0031$ and $\Omega = 0.000507$ with the physical box sizes scaled
accordingly.\footnote{Care must be taken so that the rescaling does not
correspond to a simple shift by an integer number of bits in the
machine registers.}  An identical sinusoidal velocity perturbation was
applied to the initial state in each box.  In principle, since the
length to time ratio is preserved between these different shearing box
systems, the resulting evolution should be identical.  In the
computational experiment, however, these scaled systems evolved very
differently.

\begin{figure}
\centerline{
\psfig{file=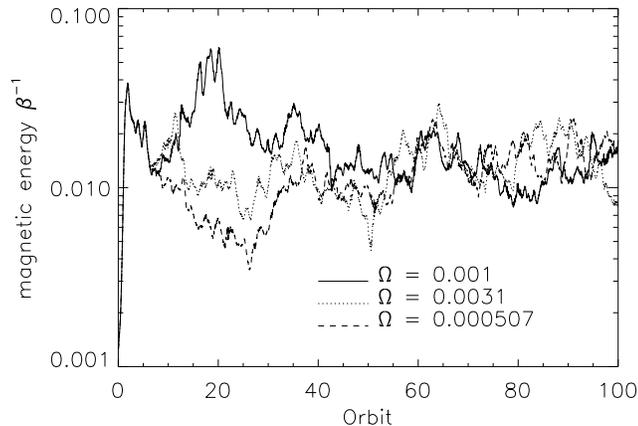,width=3.5in}}
\caption{Time evolution of the magnetic energy for a
series of simulations with different $\Omega$ values, with the box
dimensions adjusted so that each run is formally equivalent.  
Despite this, the magnetic energies diverge after the
initial linear growth and saturation.
\label{mhdo}}
\end{figure}

\begin{figure}
\centerline{
\psfig{file=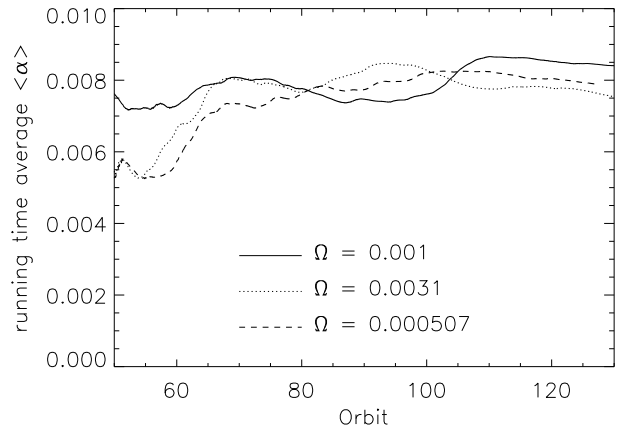,width=3.5in}}
\caption{Running time averages of $\alpha$ for the three simulations with
different $\Omega$ values.  The averages are computed beginning with
orbit 45, well after turbulence is established.
This shows that the average $\alpha$ values only slowly converge to
comparable time-averages even after 120 orbits. 
\label{ravea}}
\end{figure}

Figure \ref{mhdo} shows the volume averaged magnetic energy density
histories for the three runs.  The initial linear growth and saturation
phase is indistinguishable.  Significant deviations appear after orbit
8, and at orbit 20 the energies are widely different.  Since these
simulations correspond to identical physical conditions, one might
expect that, at the very least, time-averaged values would coincide.
Figure \ref{ravea} illustrates how the running time averages of the
volume-averaged 
stress parameter $\alpha$ for each run evolves with time.  The running
time average is defined
\begin{equation}
\langle \alpha(t)\rangle = {1\over (t-t_o)} \int_{t_o}^t \alpha dt,
\end{equation}
and the expectation is that this should converge to a steady value
characteristic of the turbulence.  The question is, over what time
interval should this happen?  As figure \ref{ravea} shows, the
fluctuations in the stress 
are so large that the running time averages $\langle \alpha(t)\rangle$ do
{\em not} coincide.   Since a very large excursion in magnetic energy is
visible in figure \ref{mhdo} at 20 orbits, we computed the running
average from 45 to 120 orbits.  But over this interval a unique
characteristic $\alpha$ does not emerge even for a single simulation. 

\begin{figure}
\centerline{
\psfig{file=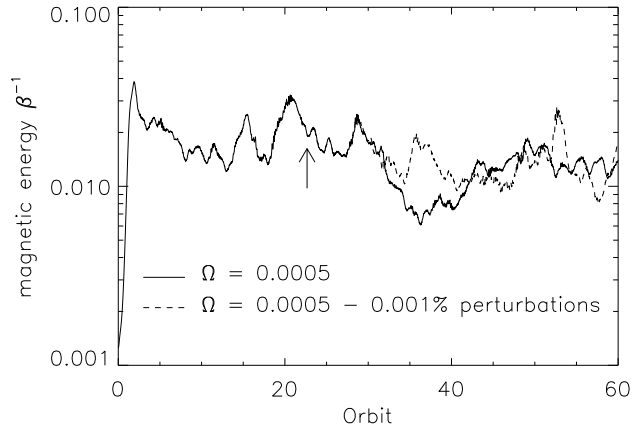,width=3.5in}}
\caption{The evolution of a baseline simulation (solid line) and a
comparison simulation to which small velocity perturbations are added
(dashed line).  Data from the baseline simulation are perturbed at the
0.001\% level at the point in time indicated by the arrow.  The
resulting magnetic energy histories diverge visibly by 30 orbits, and
are markedly different by 35 orbits.
\label{mhdpp}}
\end{figure}

The next experiment further demonstrates the chaotic nature of the
turbulence by taking the data halfway through an extended  run and
randomly perturbing the flow velocities with a gaussian rms value of
0.001\%.  The subsequent evolution is then compared with the extended run.
The parameters for this test were $\Omega$ = 0.0005, $q = 1.5$, $\beta
= 400$, $\gamma = 1.0$, $B_{z} \propto \sin (x/L_x)$, grid size of
$32 \times 64 \times 32$, and an appropriately scaled box equivalent
to the standard box.  The results are shown in figure \ref{mhdpp}.
The arrow indicates the time when the perturbations were added,
and the perturbed simulation was then run on to 60 orbits in time.
The plot of magnetic energy reveals a visible divergence by orbit 30,
and a substantial difference by orbit 35.  This is the defining feature
of chaos: substantially different macroscopic behavior created by
infinitesimal perturbations.

\begin{figure}
\centerline{
\psfig{file=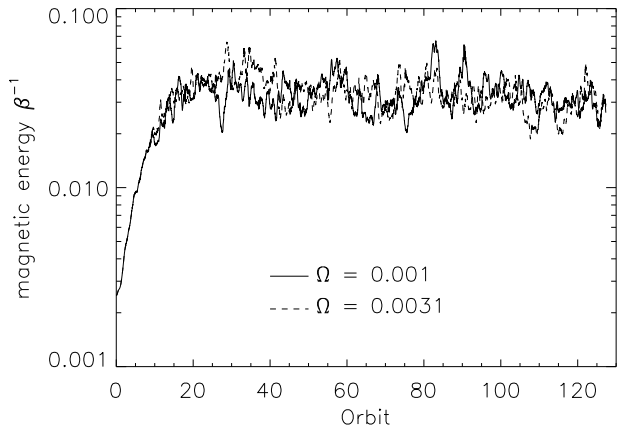,width=3.5in}}
\caption{Volume averaged magnetic energy as a
function of time for two different $\Omega$ simulations with an
initial uniform toroidal $(y)$ magnetic field.
Magnetic energy is expressed in terms of the
ratio of the magnetic pressure to initial gas pressure, $\beta^{-1}$.
\label{mhdto}}
\end{figure}

\begin{figure}
\centerline{
\psfig{file=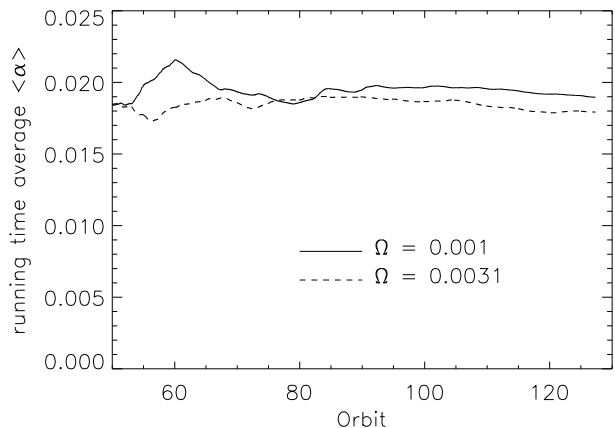,width=3.5in}}
\caption{Running time averages of $\alpha$ for two toroidal field
simulations with
different $\Omega$ values.  The averages are computed beginning with
orbit 45, well after turbulence is established.
The time averages approach different values by 120 orbits.
\label{raveat}}
\end{figure}

To compare the behavior of models with different initial field
topologies, we perform an
experiment with an initial uniform toroidal magnetic
field for two boxes, one with $\Omega=0.001$ the other with $\Omega =
0.0031$.  Other physical parameters were the same as in the vertical
field experiments.  Figure \ref{mhdto} displays the volume-averaged
magnetic energy density histories for these runs.  Again, the initial
linear growth phases of the two runs were identical; diverging behavior
is a nonlinear phenomenon.  The amplitude of the variations is not as
great as in the vertical field case.  Nevertheless, as figure
\ref{raveat} shows, the running time averages of $\alpha$ required 100
orbits to converge.

As a contrast to the chaotic behavior of MHD turbulence, we also ran
several purely hydrodynamic shearing box systems.  Hydrodynamic shearing
boxes are linearly stable when $q< 2$ (the Rayleigh criterion), and
nonlinearly stable when $q < 2-\epsilon$, where $\epsilon$ is a number on
order $0.01$ \citep{hbw99}.  All of the $q=1.5$ hydrodynamical simulations
gave similar results.  The volume-averaged kinetic energies die out in
an identical fashion for equivalent problems that use different $\Omega$
values.  Small fluctuations exist in the form of waves, and are typical
of decaying hydrodynamic turbulence.  These waves were reproduced almost
identically from simulation to simulation---the measured differences were
consistent with round-off error.  Additional experiments with different
$q$ parameters (up to $q = 1.95$) were carried out, and all of these
stable hydrodynamical experiments remained identical in their macroscopic
properties.   The numerical differences were small, not increasing,
and consistent with machine precision.  These results are consistent
with the conclusion that the  observed chaos is indeed a property of
the MHD turbulence, not of the shearing box itself or the numerics.

\subsection{Lyapunov Exponents}

The experiments provide compelling, but qualitative evidence that
MRI-driven MHD turbulence is chaotic.  A more quantitative
parameterization of chaotic behavior is afforded by the use of Lyapunov
exponents.  A system with $N$ degrees of freedom has $N$ Lyapunov
exponents.  The largest positive exponent represents the average rate
of divergence of two initially close evolution paths in phase space.
The evolution paths are traced out by monitoring the state vector of
the system, which we define as
\begin{equation}\label{state}
\mathbf{v} = \left(v_{x},v_{y},v_{z},\frac{B_{x}}{\sqrt{4 \pi
\rho}},\frac{B_{y}}{\sqrt{4 \pi \rho}},\frac{B_{z}}{\sqrt{4 \pi
\rho}}\right).
\end{equation}
The density itself is generally of secondary importance, and therefore
is tracked only indirectly in the Alfv\'en velocity.
Close evolution paths imply that there are small differences between
the state vectors for each of the two trajectories.  If an initially
small difference diverges exponentially between two state vectors, the
system is formally chaotic.  This corresponds to a positive Lyapunov
exponent.  

\cite{KB91} demonstrated chaos in MHD solar convection simulations in
three-dimensions and computed estimates of the Lyapunov exponents.  We
follow their procedure to perform a similar analysis for MRI-driven
turbulence.  We begin with an evolution in which MHD turbulence has
fully developed.  Then, a second evolution path is created by randomly
perturbing the velocity components of the first system's state vector.
The two state vector trajectories are then evolved for about 4 orbits,
and, at regular time intervals the fractional state vector difference
between the two paths was calculated:
\begin{equation}
\delta = { |\mathbf{V_p} - \mathbf{V}|\over |\mathbf{V}| }
\end{equation}
where $\mathbf{V_p}$ and $\mathbf{V}$ are the perturbed and original
state vectors.  Data from the last two orbits in the state vector
difference history are then fit to an exponential from which we
estimate the value of the largest Lyapunov exponent.

In the first experiment we employ a standard shearing box, namely $1
\times 2\pi \times 1$, grid size of $32 \times 64 \times 32$, with an
isothermal equation of state, $\gamma = 1.0$, sinusoidal vertical
field, $B_{z} \propto \sin (x/L_x)$,  of average magnetic energy
$\beta = 800$, and a background Keplerian shear, $q = 1.5$.   This was
evolved until MHD turbulence had fully developed.  Perturbations were
then applied to a parallel computation and the state vector difference
was calculated as a function of time.  The resulting Lyapunov exponent
is $0.458 \omega_{\mathrm{max}}$, where $\omega_{\mathrm{max}} =
0.75\Omega$ is the maximum unstable growth rate for the {\em linear
phase} of magnetorotational instability \citep{bh98}.  All of our
experiments yield a Lyapunov exponent comparable to
$\omega_{\mathrm{max}}$.

\begin{figure}
\centerline{
\psfig{file=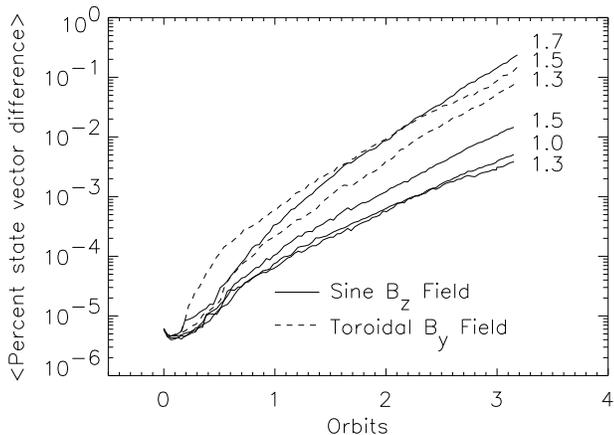,width=3.5in}}
\caption{The percent state vector difference histories
for a series of vertical field simulations (solid lines) and toroidal
field simulations (dashed lines).  Each curve is labeled by the
background shear parameter $q$.  The
exponential divergence that sets the first Lyapunov exponent is clearly
seen.  The rough trend of larger Lyapunov exponent with increased $q$ value
is also visible.
\label{e2plot}}
\end{figure}

That the Lyapunov exponent would be on order of the maximum MRI
growth rate is not surprising.  It is precisely the linearly unstable,
exponentially growing MRI that is feeding the turbulence, and driving
exponential divergence of the state vector.  To examine this more
carefully,  a series of experiments was performed on shearing boxes with
a variety of background shear $q$ parameters and for both sinusoidal
vertical and toroidal initial field configurations.  Because the
maximum linear growth rate of the MRI is $q/2$ \citep{bh98}, the
ensemble of simulations spans an interesting range of MRI growth rates.
We would expect the Lyapunov exponent to be proportional to $q$ as well.
Figure \ref{e2plot} displays the volume averaged, percent state vector
difference histories for these runs.  The curves are labeled by their $q$
values, with solid lines for the vertical field runs and dashed lines for
the toroidal field runs.  The corresponding first Lyapunov exponent values
normalized by $\omega_{\mathrm{max}}$ are $0.521$, $0.458$, $0.422$ and
$0.583$ for the vertical field cases, in order of descending $q$ value,
and $0.484$ and $0.644$ for the toroidal field $q=1.5$ and $q=1.3$ runs.
Clearly, the first Lyapunov exponents are positive, and of the same
magnitude, normalized by $\omega_{\mathrm{max}}$.  It should also be
noted that first Lyapunov exponents were calculated at many points in
time in the simulations, always with similar results.

In summary, the Lyapunov exponents in MRI-driven MHD turbulence
simulations are all positive, and, when normalized by
$\omega_{\mathrm{max}}$, they all lie within in a range of 0.4 to 0.6.
There is a slight trend of larger Lyapunov exponent for larger $q$
value (non-normalized).  Previous experiments also found stronger
overall levels of turbulence with larger $q$ values \citep{hbw99}; in a
sense, a larger Lyapunov exponent is ``more turbulent.'' Finally, the
chaos parameters of the turbulence are independent of the initial
magnetic field configuration.  Initial vertical magnetic fields have
similar Lyapunov exponent values as initial toroidal magnetic
fields.

\section{Conclusions}

Shearing box simulations of the MRI constitute an excellent numerical
laboratory in which to study chaos in turbulent flow.  Their
compactness and simple boundary conditions make them a very convenient
system to study, but they also require care to interpret.  For example,
the variance of the the flow fluctuations, which may be of direct
astrophysical interest because of its connection with radiative
emission, is a function of the box size adopted (cf. figure \ref{ziso}).

In this paper, we have demonstrated the extreme sensitivity of MHD
turbulence to infinitesimal deviations in the flow.  This was done by
several different methods:  showing that invariant scaling laws fail
when implemented numerically, for both vertical and toroidal initial
fields, and externally imposing tiny perturbations on an established
turbulent flow and following the growing deviations in the subsequent
evolution of the original and the perturbed system.  Estimates of the
largest Lyapunov exponent in a variety of turbulent flows with
different field geometries yielded values near the characteristic
growth rate of the linear MRI.  This is an indication that the linear
physics of the instability plays an active role in defining the highly
nonlinear turbulent dynamics of these flows.  One way this could come
about would be if the energy input into a Kolmogorov-like cascade was
essentially the linear MRI.

The most important practical consequence of this behavior is that the
nongaussian statistical properties of chaotic flows severely limit the
extent to which $\alpha$ modeling can be used uncritically.  Though
Maxwell and viscous stress have some formal properties in common
\citep{bp99}, the averaging procedure necessary for a semi-local
treatment of the turbulence is a very delicate matter.  A time base of
hundreds of orbits is clearly necessary to establish a meaningful
estimate of the characteristic stress.  In astrophysical systems,
especially those in transience, it may not be possible to ascribe an
instantaneous $\alpha$ value to the stress, and there may be no
recourse other than detailed numerical modeling.

\section*{Acknowledgements}
We acknowledge support under NSF grant AST-0070979,
and NASA grant NAG5-9266.  Some of the simulations described here
were carried out on computational platforms at the NSF-supported National
Center for Supercomputing Applications and 
the San Diego Supercomputer Center.

\label{lastpage}

\begin{thebibliography}{}

\bibitem[Balbus \& Hawley (1991)]{bh91} Balbus, S.~A., \& Hawley,
J.~F.  1991, ApJ, 376, 214
\bibitem[Balbus \& Hawley (1998)]{bh98} Balbus, S.~A., \& Hawley, J.~ F.
1998, Rev. Mod. Phys., 70, 1
\bibitem[Balbus \& Papaloizou  (1999)]{bp99} Balbus, S.~A., \&
Papaloizou, J.~C.~B. 1999, ApJ, 521, 650
\bibitem[Brandenburg et al.~(1995)]{betal95} Brandenburg, A., Nordlund,
\AA., Stein, R.~F., \& Torkelsson 1995, ApJ, 446, 741
\bibitem[Hawley, Balbus, \& Winters (1999)]{hbw99} Hawley, J.~F., 
Balbus, S. A., \& Winters, W. F. 1999, ApJ, 518, 394
\bibitem[Hawley, Gammie \& Balbus (1995)]{hgb95} Hawley, J.~F.,
Gammie, C.~F., \& Balbus, S. A.  1995, ApJ, 440, 742
\bibitem[Hawley, Gammie \& Balbus (1996)]{hgb96} Hawley, J.~F.,
Gammie, C.~F., \& Balbus, S. A.  1996, ApJ, 464, 690
\bibitem[Hawley \& Stone (1995)]{hs95}Hawley, J.~F., \& Stone, J.~M.
1995, Comp Phys Comm, 89, 127
\bibitem[Kurths \& Brandenburg (1991)]{KB91} Kurths, J., \& Brandenburg,
A., 1991, Phys Rev A, 44, R3427
\bibitem[Shakura \& Sunyaev (1973)]{ss73} Shakura, N.~I., \& Sunyaev, R.~A.
1973, A\&A, 24, 337
\bibitem[Stone \& Norman (1992a)]{sn92a} Stone, J.~M., \& Norman, M.
L.  1992a, ApJS, 80, 753
\bibitem[Stone \& Norman (1992b)]{sn92b} Stone, J.~M., \& Norman, M.
L.  1992b, ApJS, 80, 791


\end{thebibliography}
\end{document}